\documentclass[aps,pre,twocolumn,showpacs,amsmath,amssymb,superscriptaddress]{revtex4-1}
\usepackage{epsfig}
\usepackage{graphicx}
\usepackage{dcolumn}
\usepackage{bm}
\usepackage{float}

\begin{document}

\title{Efficient algorithm to study interconnected networks}

\author{Christian M. Schneider}
\email{schnechr@mit.edu}
\affiliation{Department of Civil and Environmental Engineering, Massachusetts Institute of Technology, 77 Massachusetts Avenue, Cambridge, MA 02139, USA}
\affiliation{Computational Physics for Engineering Materials, IfB, ETH Zurich, Wolfgang-Pauli-Strasse 27, CH-8093 Zurich, Switzerland}
\author{Nuno A. M. Ara\'ujo}
\email{nuno@ethz.ch}
\affiliation{Computational Physics for Engineering Materials, IfB, ETH Zurich, Wolfgang-Pauli-Strasse 27, CH-8093 Zurich, Switzerland}
\author{Hans J. Herrmann}
\email{hans@ifb.baug.ethz.ch}
\affiliation{Computational Physics for Engineering Materials, IfB, ETH Zurich, Wolfgang-Pauli-Strasse 27, CH-8093 Zurich, Switzerland}
\affiliation{Departamento de F\'{\i}sica, Universidade Federal do Cear\'a, 60451-970 Fortaleza, Cear\'a, Brazil}

\begin{abstract}

Interconnected networks have been shown to be much more vulnerable to random
and targeted failures than isolated ones, raising several interesting
questions regarding the identification and mitigation of their risk. The
paradigm to address these questions is the percolation model, where the
resilience of the system is quantified by the dependence of the size of
the largest cluster on the number of failures. Numerically, the major
challenge is the identification of this cluster and the calculation of
its size. Here, we propose an efficient algorithm to tackle this
problem. We show that the algorithm scales as $O(N\log{N})$, where $N$
is the number of nodes in the network, a significant improvement
compared to $O(N^2)$ for a greedy algorithm, what permits studying much
larger networks. Our new strategy can be applied to any network topology
and distribution of interdependencies, as well as any sequence of
failures.

\end{abstract}

 

\maketitle

\section{Introduction}

Most real networks are strongly dependent on the functionality of other
networks \cite{Barabasi,Watts,Dorogovtsev,Caldarelli, Barrat}. For example, the performance of a power grid is assured by a
system of global monitoring and control, which depends on a
communication network. In turn, the servers of the communication
network rely on the power grid for power supply. This interdependence
between networks strongly affects their resilience to failures. Buldyrev
\textit{et al.} \cite{Buldyrev2010} have developed the first strategy to analyze
this coupling in the framework of percolation. To the conventional
representation of complex networks, where nodes are the agents (e.g.,
power stations or servers) and edges are the interactions (either
physical or virtual), they added a new type of edges, namely,
dependency links, to represent the internetwork coupling. Such links
couple two nodes from different networks in such a way that if one fails
the other cannot function either. They have shown that this coupling
promotes cascading failures and strongly affects the systemic risk,
drawing the attention towards the dynamics of coupled systems. A different
framework based on epidemic spreading has also been proposed leading to
the same conclusions \cite{Son2012}.

To quantify the resilience of interconnected networks, one typically
simulates a sequence of node failures (by removing nodes) and measures
the dependence of the size of the largest connected component on the
number of failures \cite{Schneider2011a,Schneiderpre}. The first studies
have shown that, depending on the strength of the coupling (e.g.,
fraction of dependency links), at the percolation threshold, this
function can change either smoothly (weak coupling) or abruptly (strong
coupling) \cite{Gao2012a}. As reviewed in
Refs.~\cite{Vespignani,Gao2012a,Havlin2012}, several works have followed studying,
for example, the dependence on the coupling strength
\cite{Parshani2010a,Schneiderpre}, the role of network topology, and the
phenomenon on geographically embedded networks
\cite{Son2011,Barthelemy2011}.  A more general framework was also
developed to consider a network of networks
\cite{Gao2011,Xu2011,Brummitt2012}.
In all cases, astonishing properties have been revealed, which were
never observed for isolated systems.

For many cases of interest, the size of the largest component needs to
be computed numerically as the available analytic formalisms are limited
to very simple networks, interdependencies, and sequence of failures
\cite{Buldyrev2010,Parshani2010a,Gao2012a}. However, the determination
of this largest component and its size is not a trivial task. When a
node is removed (fails), the triggering of cascading failures and
multiple interdependencies need to be considered.  Here we propose an
efficient algorithm, where a special data structure is used for the fast
identification of the largest fragment when the network breaks into
pieces. We show that the algorithm scales as $O(N \log{N})$, where $N$
is the number of nodes in the network, while the one of a greedy
algorithm is $O(N^2)$.  This strategy permits studying very large system
sizes and many samples, which leads to much more accurate statistics.
Since our description is generic, it is possible to consider any network
and distribution of interdependencies, as well as sequences of failures
\cite{Holme2002,Schneider2011b,Schneider2012}.

The paper is organized in the following way. The algorithm is described
in Sec.~\ref{sec::algorithm} and its efficiency discussed in
Sec.~\ref{sec::comp}. In Sec.~\ref{sec::finalremarks} we make some final
remarks and discuss possible future applications.

\section{Algorithm}\label{sec::algorithm}
\begin{figure}
 \includegraphics[width=8.5cm,angle = 0]{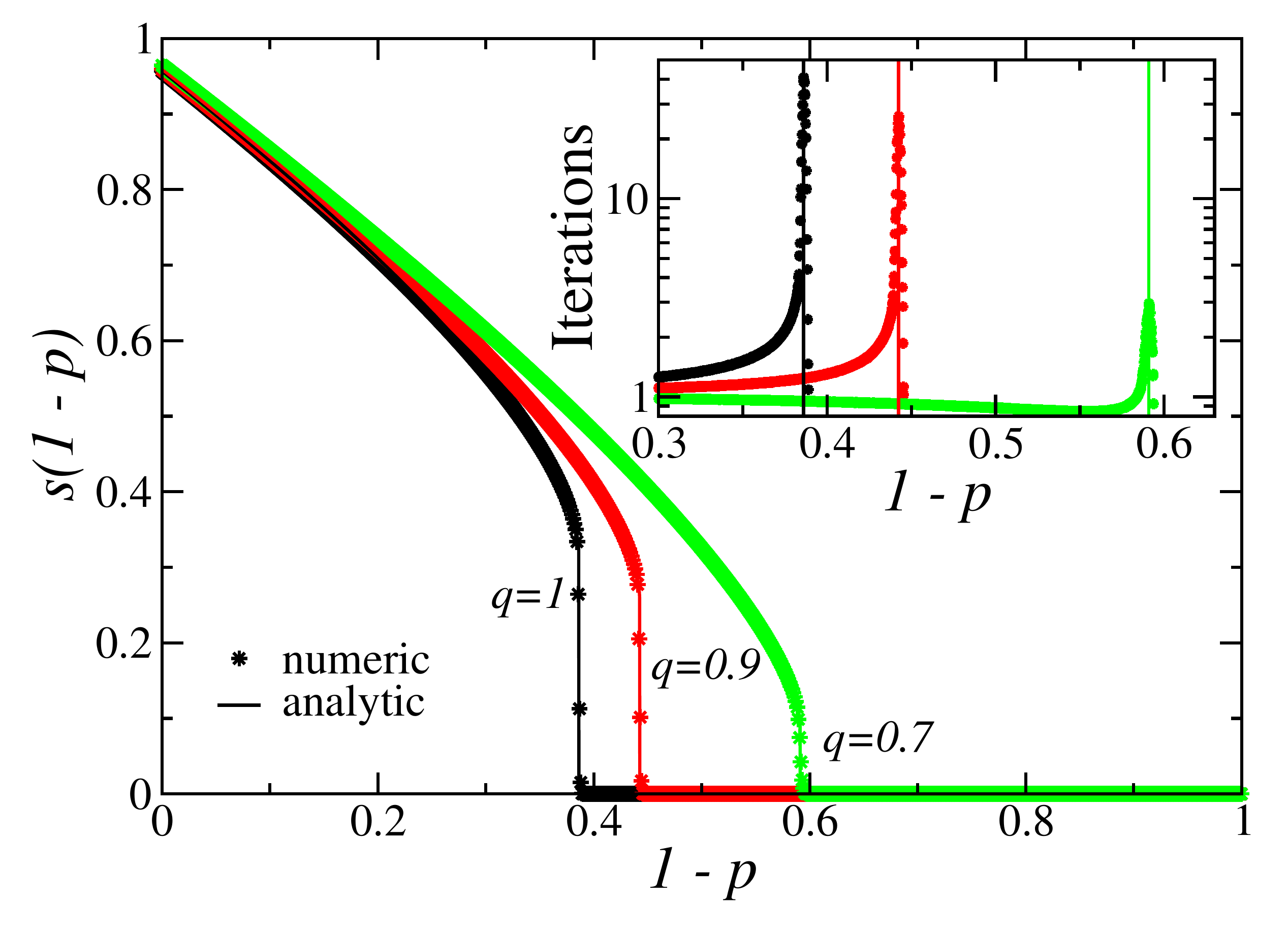}
 \caption{
 (Color online) Comparison between numerical and analytic results for
percolation on two coupled Erd\H{o}s-R\'enyi networks. In the main plot,
the numerical (data points) results for the size of the largest
connected cluster are obtained from two coupled networks with $N =
512000$ nodes each and coupling strengths of \mbox{$q = 1, 0.9,$ and
$0.7$} (from left to right). The lines correspond to the analytic
results computed as in Ref.~\cite{Parshani2011}. The numerical results
are averages over $50$ different pairs of networks and $100$ sequences
of random failures. All numerical and corresponding analytic curves
overlap. In the inset we see the number of iterations per failure, defined
as the number of times a cascade triggered by a node failure propagates
back and forth between the two networks. The lines tag the analytic
results for the percolation threshold.
Initially, a node removal does not trigger any cascade but, as one
approaches the percolation threshold, very large cascades occur,
resulting in the collapse of the entire system.
  \label{fig:0} }
\end{figure}

Figure~\ref{fig:0} shows the dependence of the fraction of nodes in the
largest connected cluster $s$ on the fraction of removed nodes $1-p$,
for two Erd\H{o}s-R\'enyi networks with more than one million nodes. When
nodes are randomly removed, the largest connected component decreases in
size, until the network is completely fragmented above a threshold
$1-p_c$. In the inset, we see the evolution of the number of iterations
per removed node. An iteration corresponds to a set of failures in one
network triggered by an internetwork coupling, i.e., by the removal of a
dependency link. The number of iterations is negligibly small for low
values of $1-p$ but peaks at the threshold. Following the cascade after
removing a node is the most computational demanding task. Consequently,
an efficient algorithm is required to identify, in a fast way, if a
node removal triggers a cascade or not.

\begin{figure}
 \includegraphics[height=3.0cm,angle = 0]{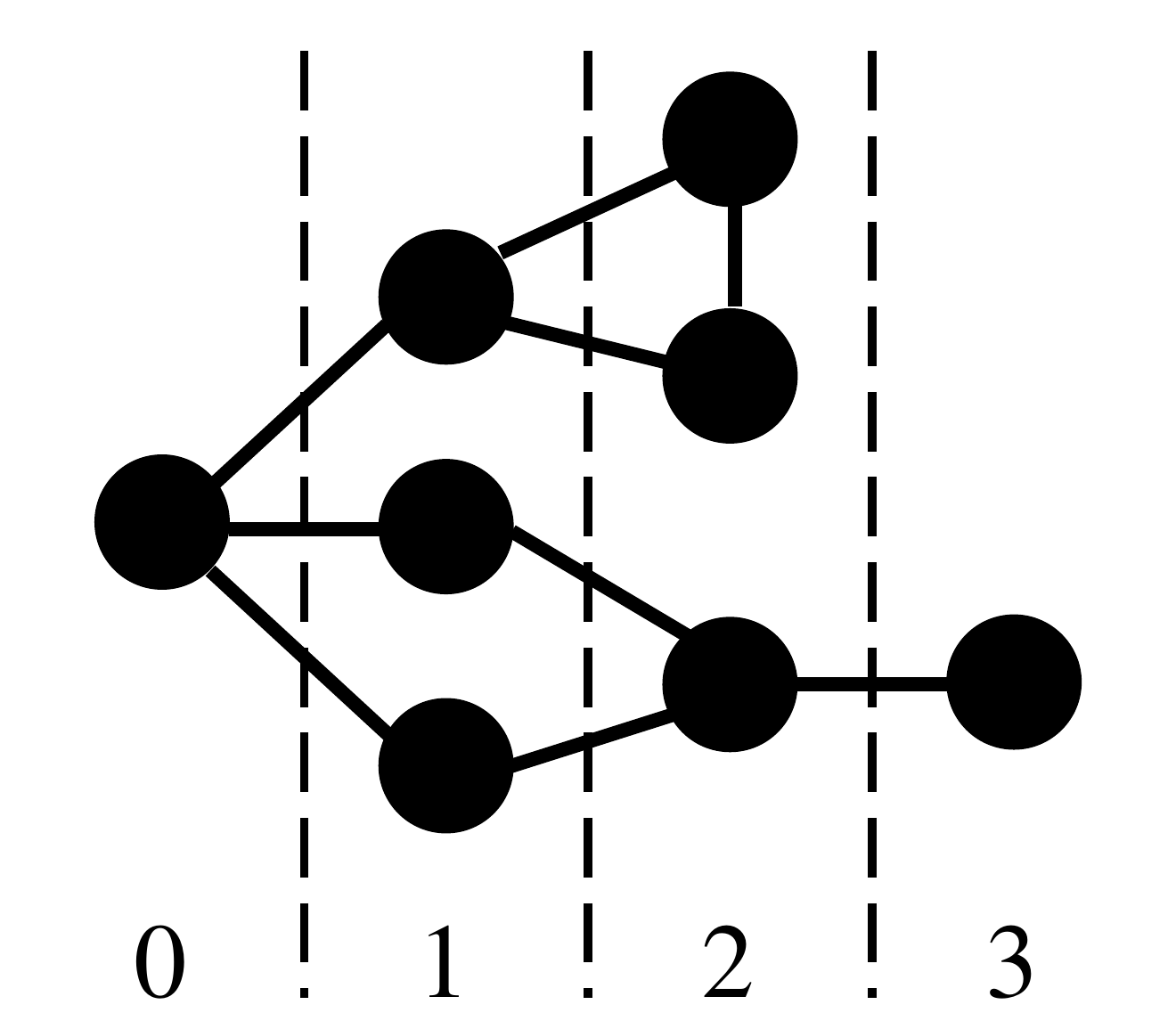}
 \caption{
 Example of the level structure for a small network. An energy number is
assigned to each node, which corresponds to the shortest distance to the
root node (the most central node in this example).
  }
\label{fig:1}
\end{figure}

Here, we propose an efficient data structure to recognize the beginning
of a cascade and identify the different fragments resulting from a node
removal. Since we are interested in the evolution of the largest
connected component, we only follow this cluster. 
Our algorithm uses a hierarchical data structure
with different levels. As illustrated in Fig.~\ref{fig:1}, we choose the
node with the highest degree as the root and assign to it the level $L =
0$. All neighbors of this root are on the second level ($L = 1$) and
they are directly connected to the root. All neighbors of the second
level, which have not an assigned level yet, are then placed on
the third level. We proceed iteratively in the same way, until all nodes
of the cluster have a level. Note that we can have links within the same
level and between levels but, in the latter, the level difference is
limited to unity. The depth of the level structure is the maximal
distance between the root and any other node in the network. For random
networks, this depth approximately scales with $\log{N}$
\cite{Albert2002,Newman2003} and it scales even slower for many
scale-free networks \cite{Cohen2003}. Note that, in the case of $n$
coupled networks we will have $n$ different hierarchical structures, i.e., one per
network, representing its largest component.

\begin{figure}
 \includegraphics[height=3.0cm,angle = 0]{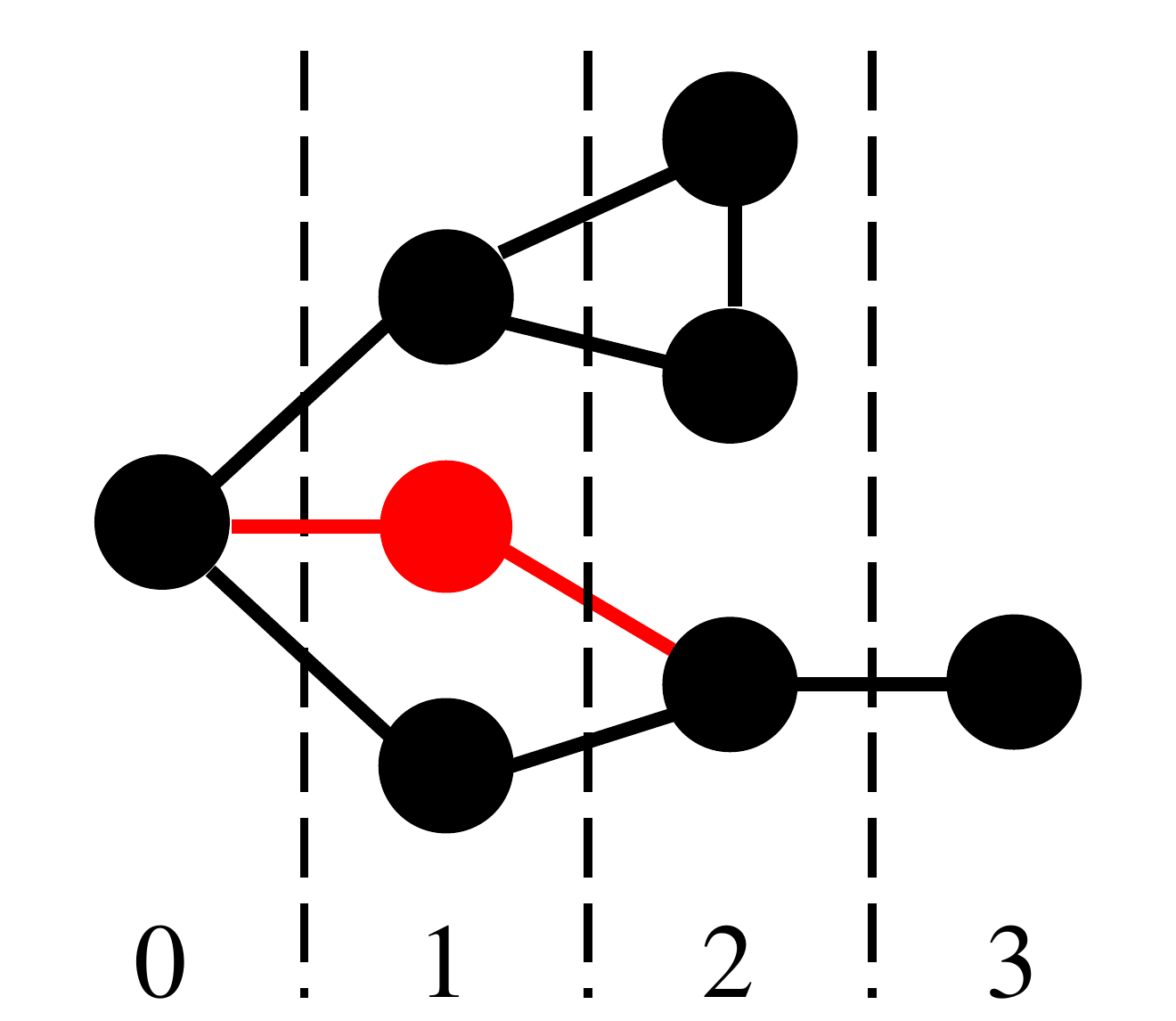}
 \includegraphics[height=3.0cm,angle = 0]{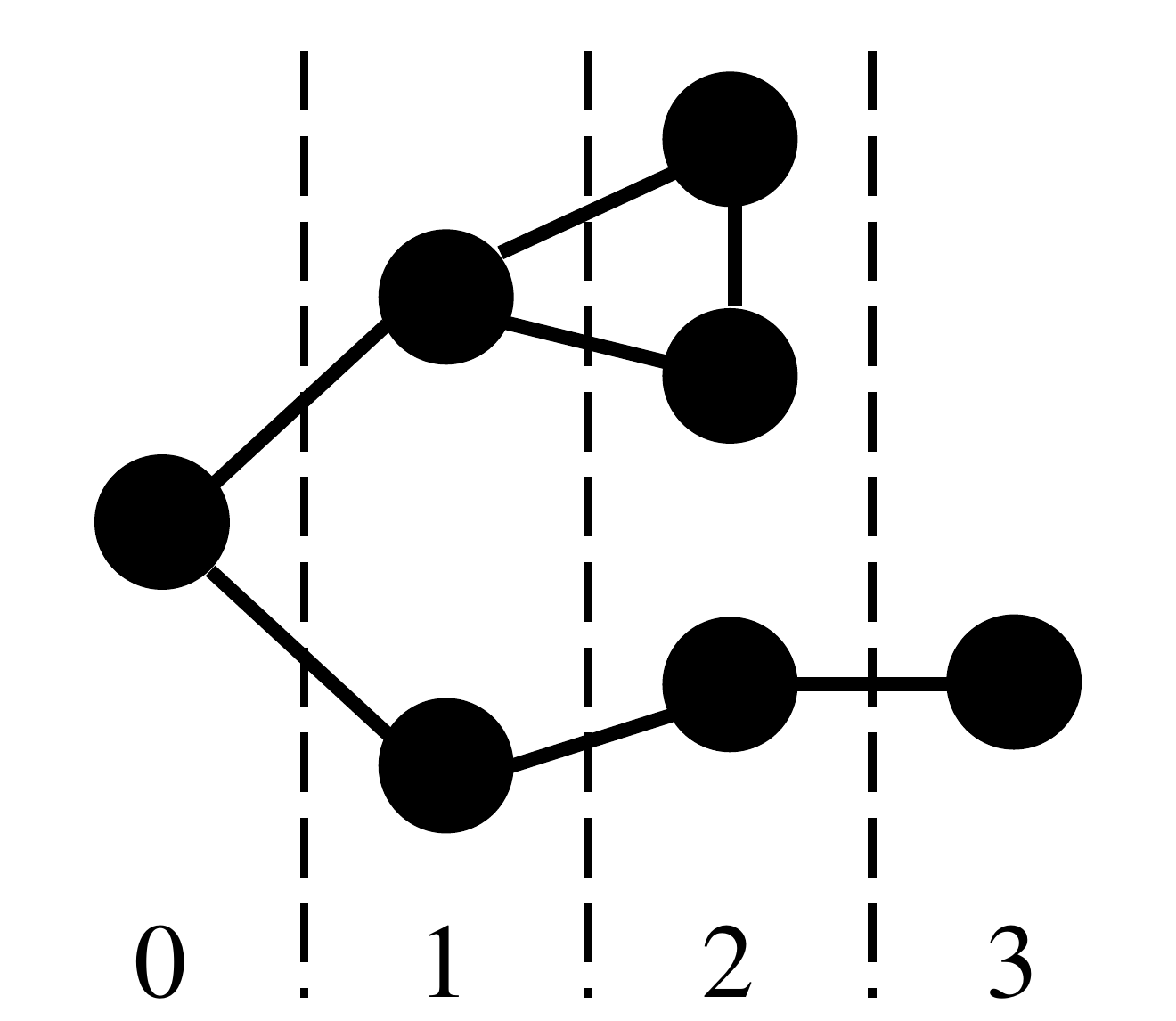}
 \caption{
 (Color online) Example of a node removal. In this case, the red (light) node with $L =
1$ is removed. Since the only neighbor of this node with higher level
has another neighbor with $L = 1$, the size of the largest connected
cluster is only reduced by one.
  }
\label{fig:2}
\end{figure}

\begin{figure}
 \includegraphics[height=3.0cm,angle = 0]{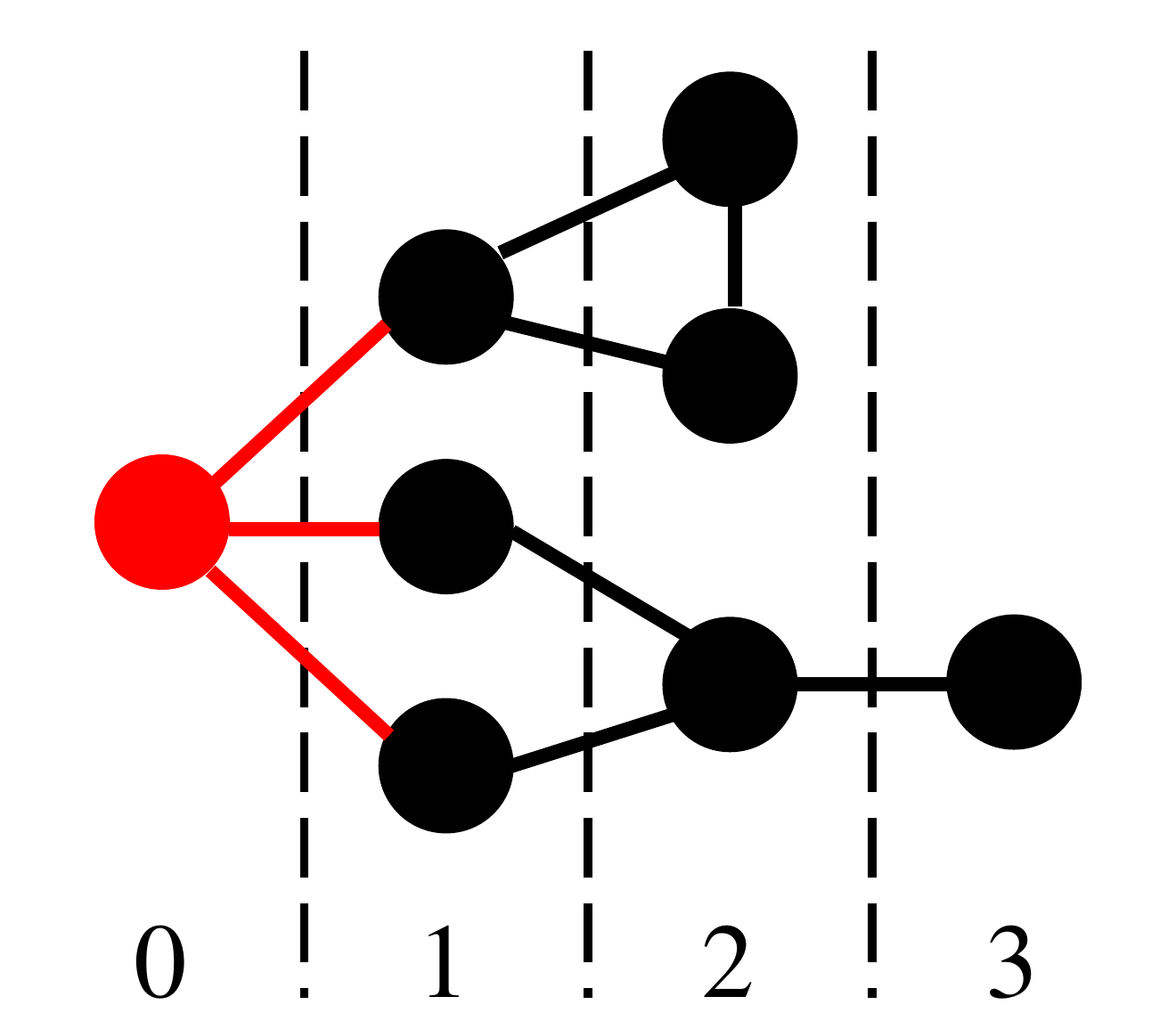}
 \includegraphics[height=3.0cm,angle = 0]{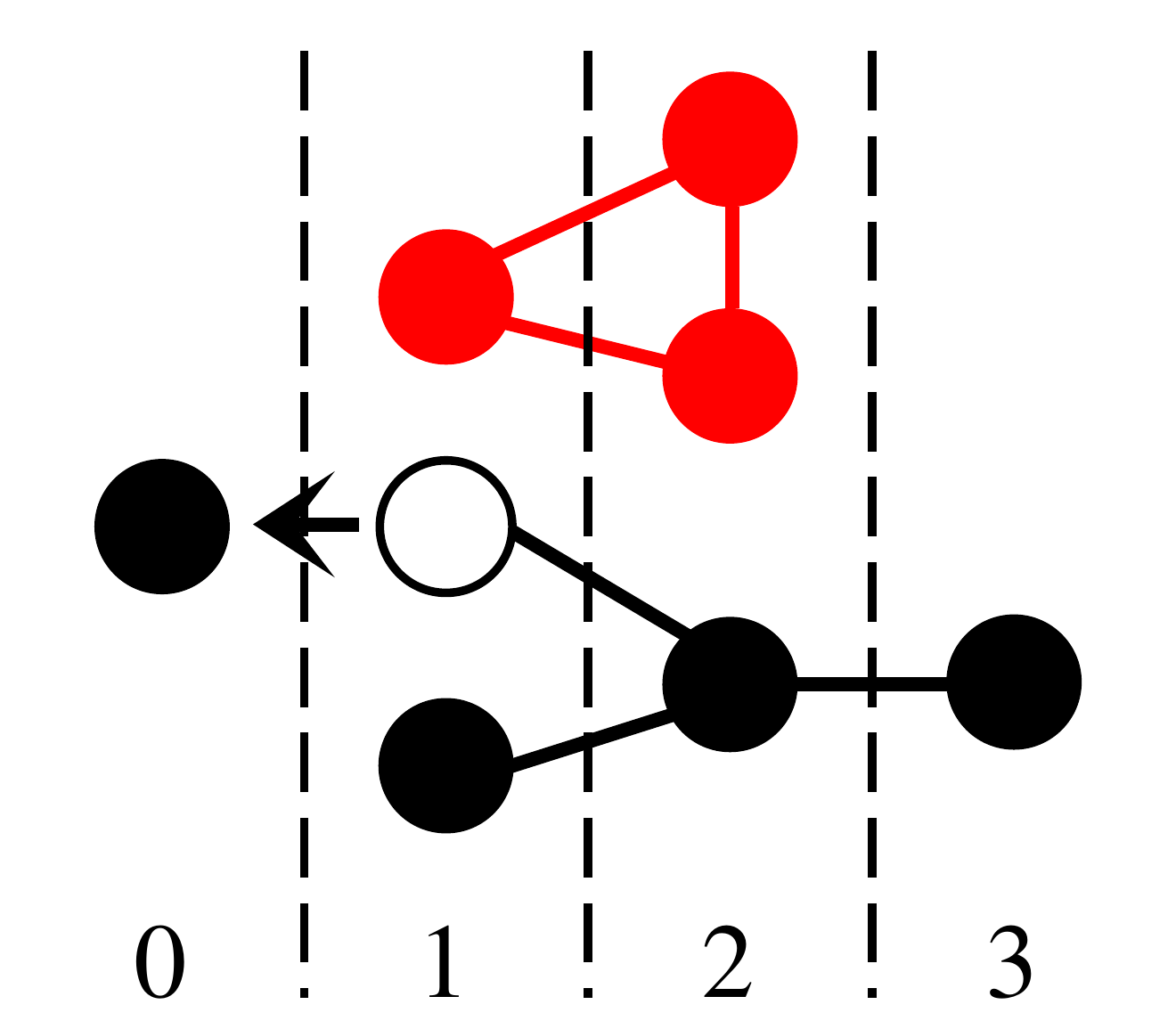}
 \includegraphics[height=3.0cm,angle = 0]{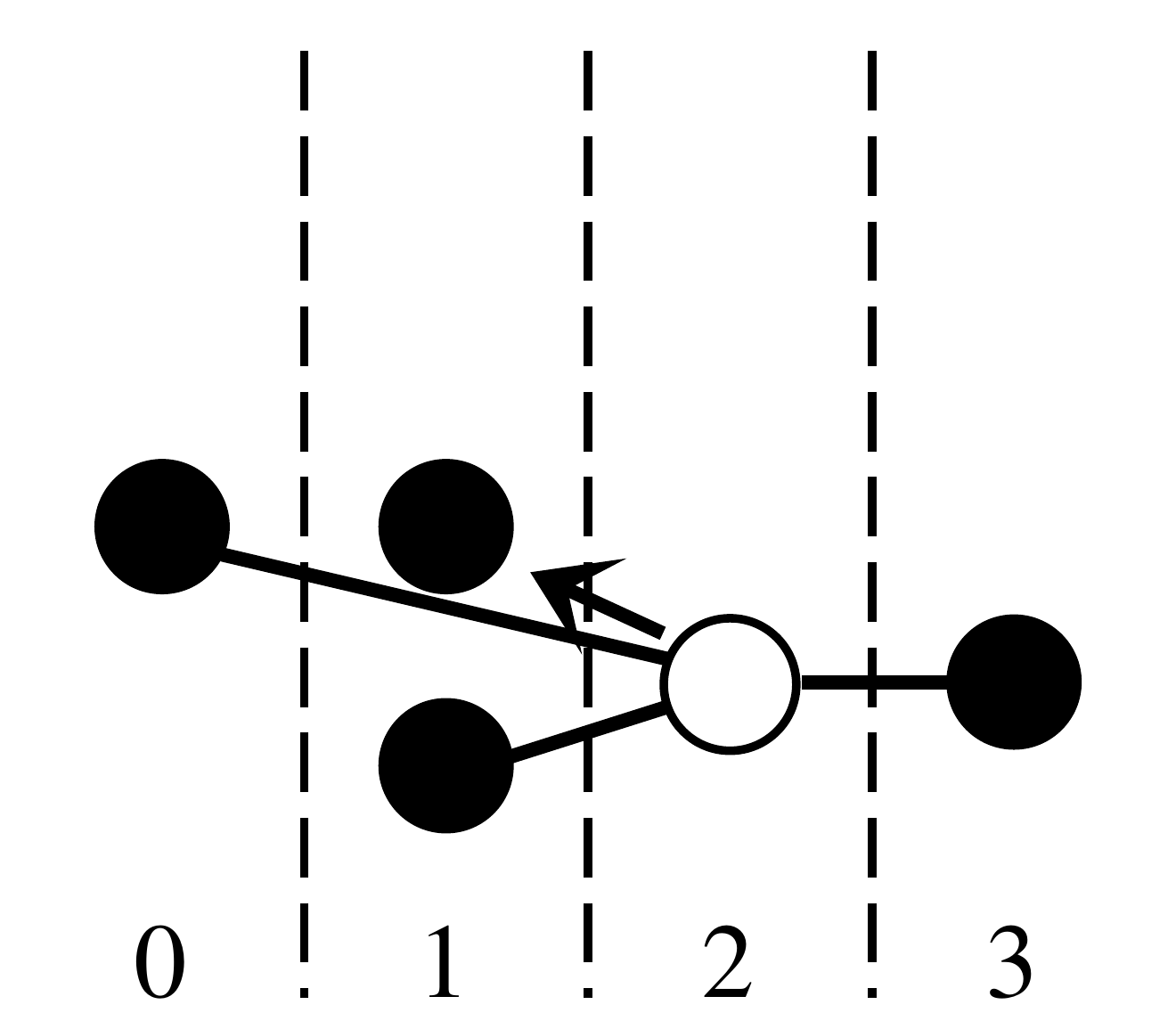}
 \includegraphics[height=3.0cm,angle = 0]{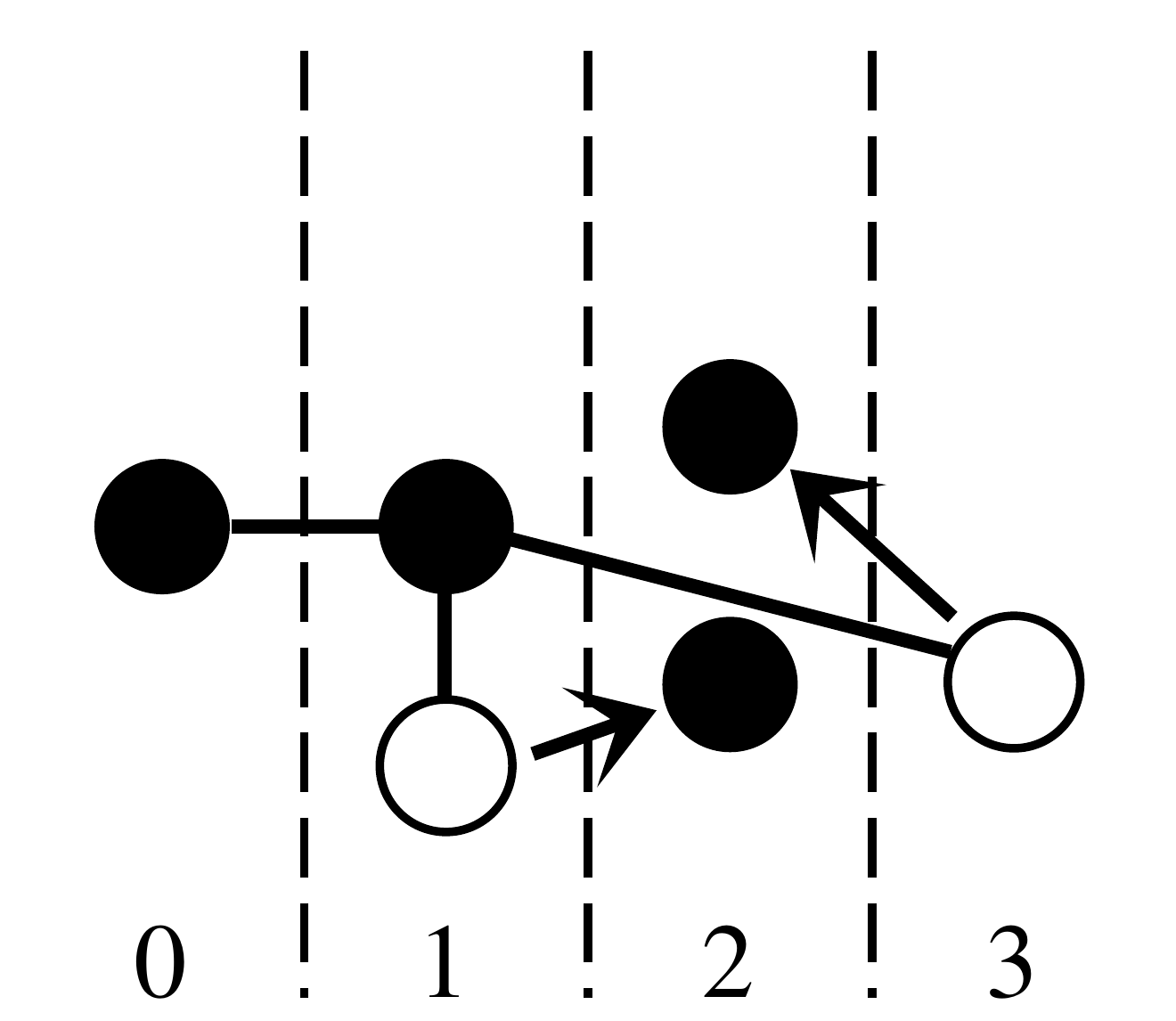}
 \includegraphics[height=3.0cm,angle = 0]{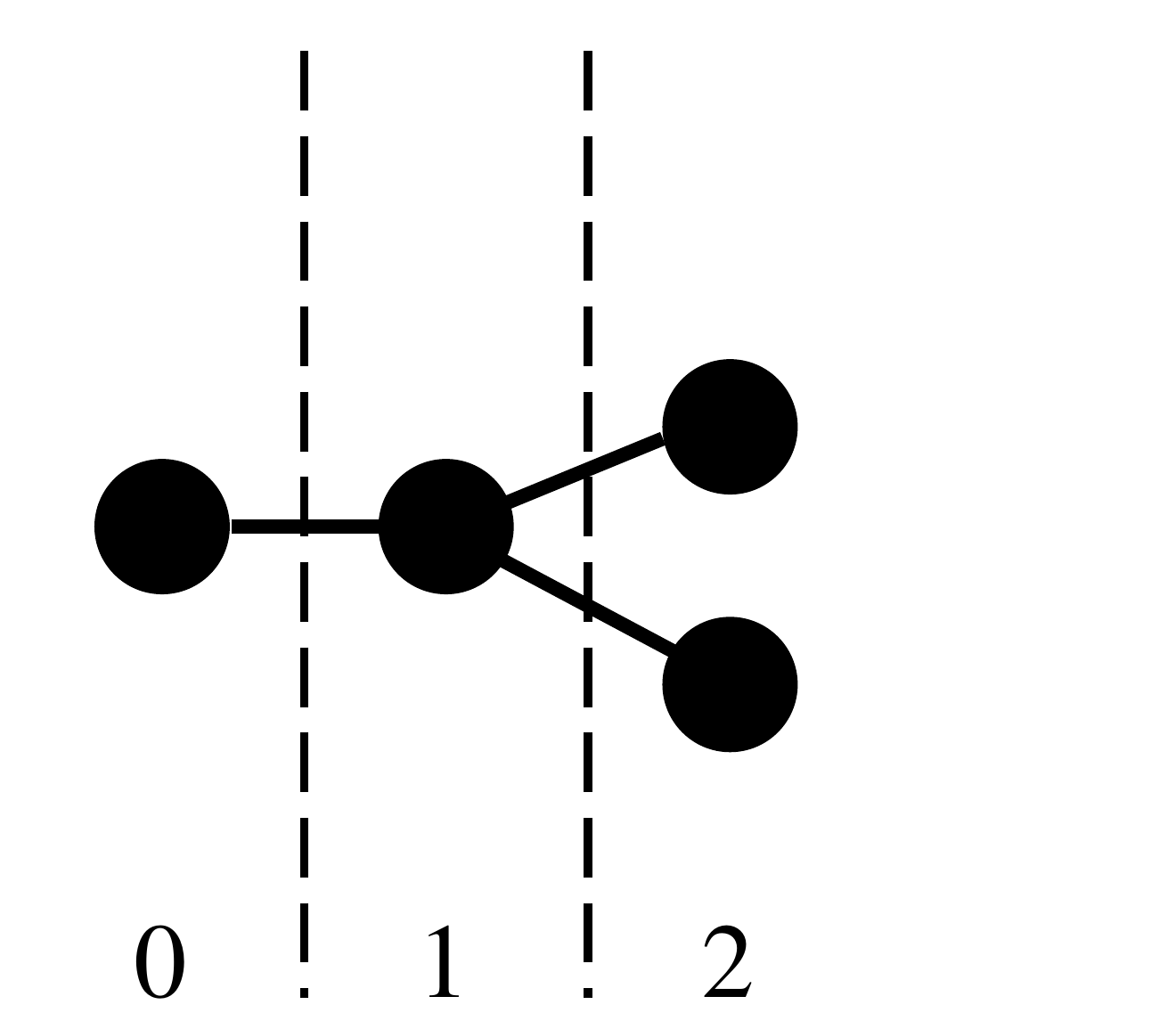}
 \caption{
 (Color online) Example of a reorganization due to a node removal.
In the example, the root node is removed, thus the level structure has to be
reorganized. First, the largest subnetwork is identified and all the
other subnetworks are removed. One of the surviving first neighbors of
the old root is randomly selected to become the new root. Then the
levels of its neighbors are updated as well as their branches. Note that
the update of a branch is complete when the level of a node remains the
same. In the worst case scenario, the complexity of the entire update
process is $O(M)$, where $M$ is the size of the entire branch.
 }
\label{fig:4}
\end{figure}

When a node in level $L$ is removed, the ordering needs to be updated.
All neighbors at a higher level $L+1$ which are connected to another
node in level $L$ remain in the same level, as shown in
Fig.~\ref{fig:2}. The nodes in level $L+1$ which have no further
neighbors in level $L$ but only in level $L+1$, need to be updated
(moved one level up) as well as the entire branch connected to them. In
those two cases, the size of the largest connected component
in this iteration is just changed by unity (the initially removed node).
If neither of those cases occurs, i.e. all neighbors have a higher level, we
proceed iteratively through the branch of neighbors with a breadth first
search (up in level) until we detect one node in level $L'$ which has at
least one neighbor in level $L'$ or $L' - 1$ which is not detected by
the breadth first search. In this case, the entire branch of detected
nodes is updated, starting from the last node in level $L'$. On the
other hand, if no node in the branch establishes a connection with the
other branches, it implies that the largest component was split into
subnetworks and one has to decide which one is the largest. Then the
size of the largest connected component is adjusted and all nodes
reorganized (see example in Fig.~\ref{fig:4}).

\section{Number of commands and computational time}\label{sec::comp}

\begin{figure}
 \includegraphics[width=8.5cm,angle = 0]{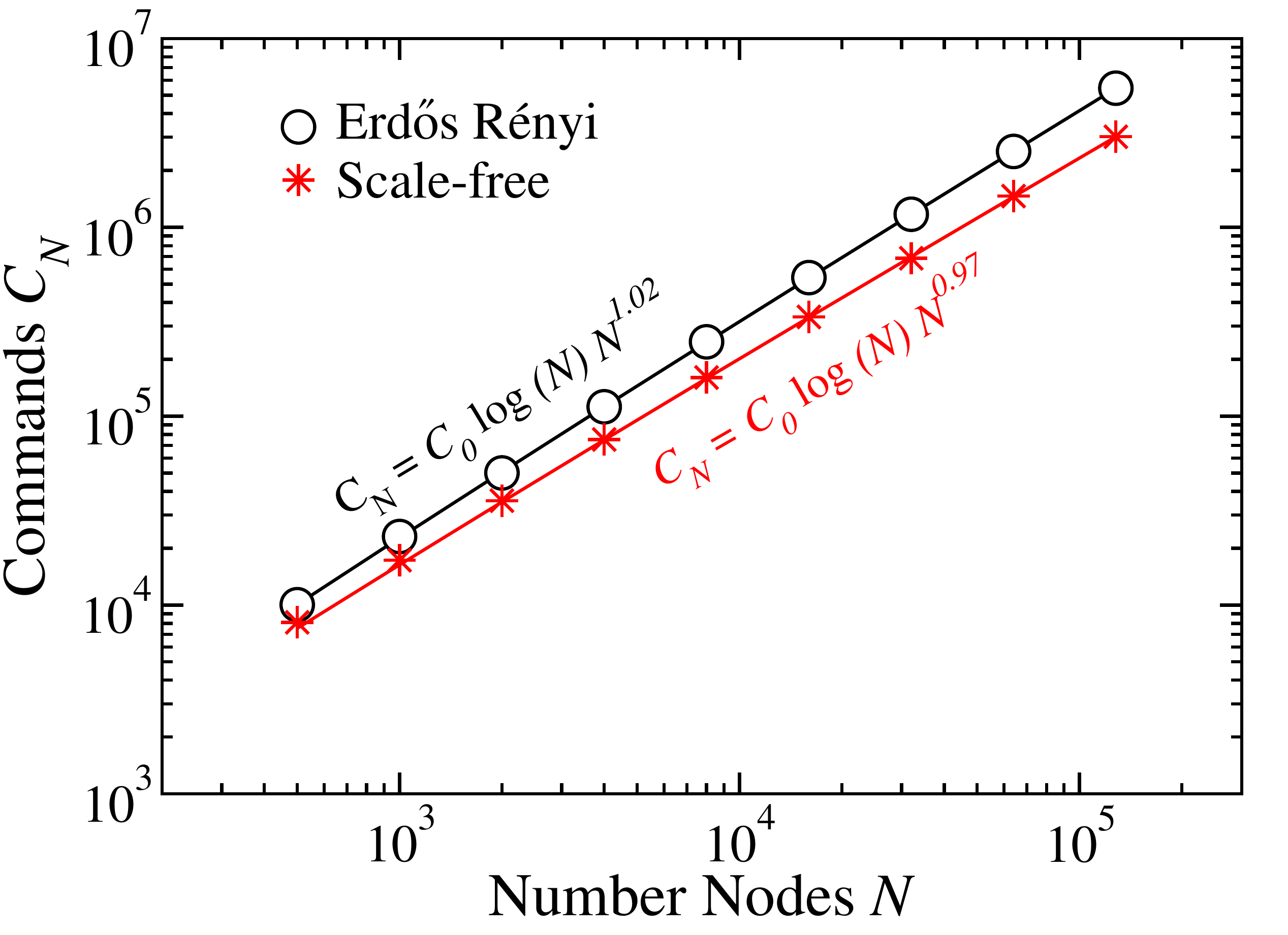}
 \caption{
 (Color online) The number of used commands in the program $C_N$ versus the system size
$N$. The function $f(N) = C_0 \log(N) N^{C_1}$ is fitted to the observed
values. The fitting parameter $C_1$ is $1.02 \pm 0.03$ for
Erd\H{o}s-R\'enyi networks and $0.97 \pm 0.04$ for scale-free networks,
respectively.}
 \label{fig:commands}
\end{figure}

To assess the efficiency of the algorithm, we study the dependence of
the number of commands $C_N$ on the network size $N$. We count as a
command, every time a node in one of the networks is removed, its level
changed, or just checked during the reorganization of the level
structure.  Figure~\ref{fig:commands} (main plot) shows the size
dependence of the average $C_N$ for two coupled Erd\H{o}s-R\'enyi networks
\cite{Erdos1960}.  The line fitting the data points is
$C_0\log(N)N^{C_1}$, where $C_1$ is $1.02\pm0.03$. In a greedy algorithm
where the largest connected cluster is recalculated by counting all
remaining nodes in this component after each node removal, the number of
commands is expected to scale as $O(N^2)$.  With our data structure,
this limit where all nodes are checked, would correspond to the worst
case scenario, where the removed nodes would systematically be the root.
Therefore, our algorithm represents a significant improvement over the
traditional greedy algorithm. Also in Fig.~\ref{fig:commands}, we plot
the number of commands $C_N$ for two coupled scale-free networks, with
degree exponent $\gamma= 2.5$ \cite{Molloy1995}. The same scaling with
the network size was found, with $C_1=0.97\pm0.04$.  So, for the two
considered types of networks, our algorithm scales with $O(N\log{N})$.
In general, this scaling will depend on how the average shortest path
scales with the network size.

\begin{figure}
 \includegraphics[width=8.5cm,angle = 0]{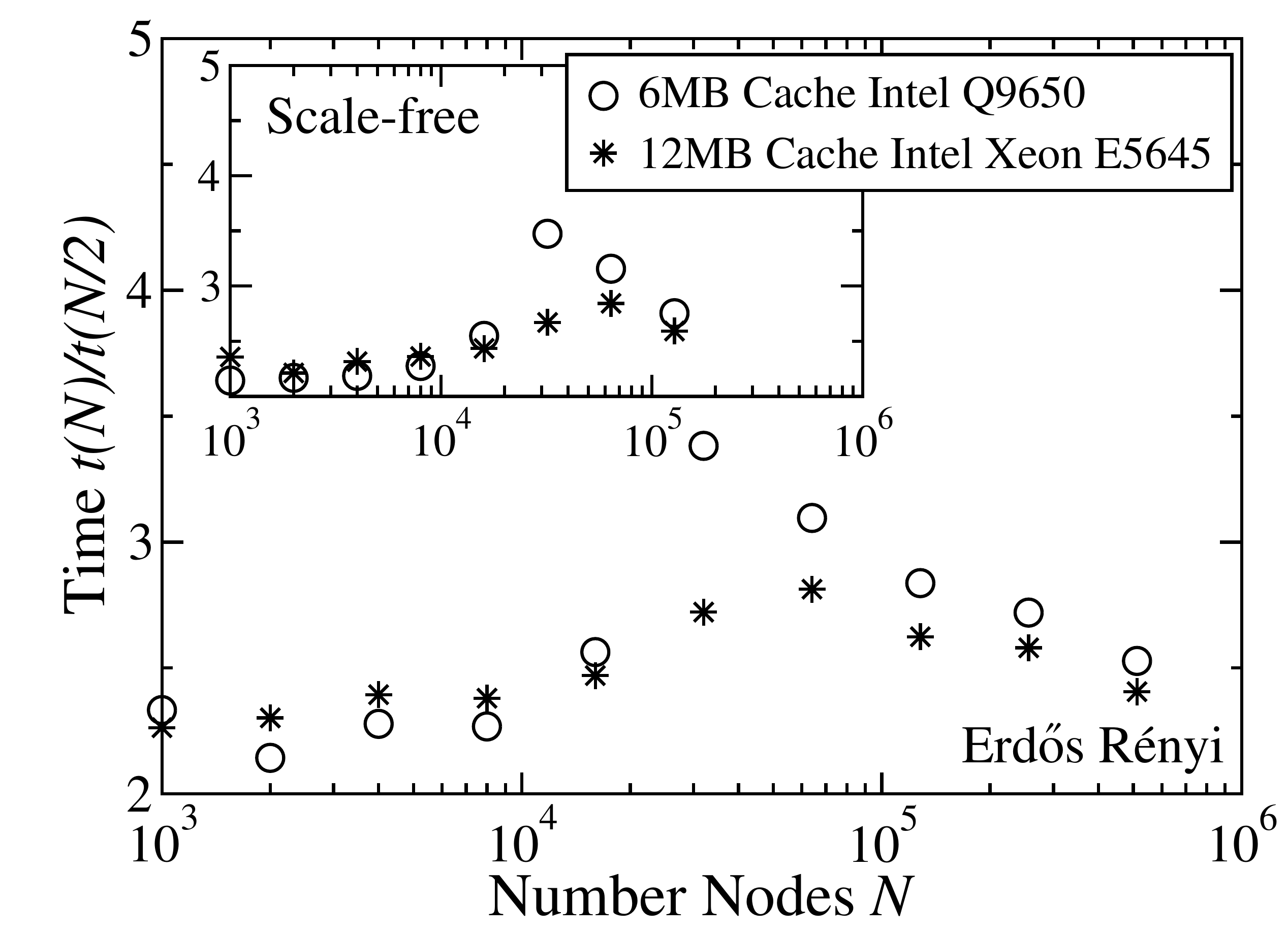}
 \caption{
  System-size dependence of the average CPU time $t(N)$, in seconds,
necessary to calculate one sequence of node removals. We see the ratio
between two times $t(N)/t(N/2)$ for two different machines (different
capacity of the CPU cache memory) for two coupled Erd\H{o}s-R\'enyi (ER,
main plot) and scale-free networks (SF, inset). Results are averages
over $100$ sequences of random removals and $50$ different networks.
}
 \label{fig:time}
\end{figure}

Figure~\ref{fig:time} shows the size dependency of the average
computational time $t(N)$ required to compute an entire sequence of node
removals. We show the ratio $t(N)/t(N/2)$ obtained from two computers
with $6$ MB and $12$ MB CPU cache for Erd\H{o}s-R\'enyi networks (main
plot) and scale-free networks (inset). In both cases, we observe a
crossover between two different scaling regimes at a certain system size
$N^*$. This crossover at $N^* = 4000$ and $N^* = 8000$ for $6$ MB cache
and $12$ MB cache, respectively, depends on the size of the CPU cache
memory (L$2$). For network sizes $N<N^*$, the size of the system is such
that all information can be kept inside the CPU cache, being more
efficient. For $N>N^*$, not all information fits in the CPU cache and
the efficiency decreases, since the access to the Random Access Memory
(RAM) is slower. 

In the first regime $N<N^*$ the increase of the CPU time is consistent
with an algorithm scaling as $O(N\log N)$. In the second regime $N \gg
N^*$ the CPU time seems to converge to the same logarithmic scaling.

\section{Final remarks}\label{sec::finalremarks}

We have proposed an efficient algorithm to monitor the size of the
largest connected component during a sequence of node failures, in a
system of interdependent networks. Although, in general, the algorithm
can be considered to study percolation in both isolated and coupled networks, its is
tailored for coupled ones. We have shown that the algorithm complexity
is $O(N \log N)$, a significant improvement over the greedy algorithm
of complexity $O(N^2)$.

With our efficient algorithm, it is now possible to simulate much larger
system sizes, a relevant feature to develop accurate studies. One of the
most striking results of coupled networks is that, for strong coupling,
the fragmentation of the network into pieces occurs in a discontinuous
way \cite{Buldyrev2010}. The possibility of accurate measurements, with
reduced finite-size effects, permits to determine with high precision the
critical coupling above which the percolation transition is
discontinuous \cite{Parshani2010a}. Our algorithm can now be applied to
any network topology, sequence of failures (e.g., random or high-degree
node) \cite{Schneiderpre}, and distribution of dependency links
\cite{Li2012}, helping clarifying how to mitigate the systemic risk
stemming from interdependencies.

\section{Acknowledgment}
We acknowledge financial support from the ETH Risk Center, from the
Swiss National Science Foundation under contract 200021 126853 and by New
England UTC Year 23 grant, awards from NEC Corporation Fund, the Solomon Buchsbaum Research Fund.

\end{document}